\documentclass[12pt]{article}
\pdfoutput=1

\usepackage[utf8]{inputenc}

\usepackage{graphicx,psfrag,epsf,color}
\usepackage{amsmath,amssymb,amsfonts}
\usepackage{array}
\usepackage{cite}
\usepackage{rotating}
\usepackage{slashed, cancel}
\usepackage{scalerel,stackengine}
\usepackage{multirow}

\numberwithin{equation}{section}

\newcommand{\be}{\begin{equation}}
\newcommand{\ee}{\end{equation}}
\newcommand{\bea}{\begin{eqnarray}}
\newcommand{\eea}{\end{eqnarray}}
\newcommand{\bi}{\begin{itemize}}
\newcommand{\ei}{\end{itemize}}
\newcommand{\ben}{\begin{enumerate}}
\newcommand{\een}{\end{enumerate}}
\newcommand{\bt}{\begin{tabular}}
\newcommand{\et}{\end{tabular}}

\newcommand{\nn}{\nonumber}

\newcommand{\nm}{n_-}
\newcommand{\np}{n_+}

\newcommand{\as}{\alpha_s}

\newcommand{\ord}{{\cal O}}

\begin{document}
\allowdisplaybreaks

\begin{titlepage}

\begin{flushright}
{\small
TUM-HEP-1235/19\\
NIKHEF/2019-048\\
CERN-TH-2019-176\\
October 28, 2019
}
\end{flushright}

\vskip1cm
\begin{center}
{\Large \bf\boldmath Leading-logarithmic threshold 
resummation \\[0.0cm]  
of Higgs production in gluon fusion \\[0.2cm]
at next-to-leading power}
\end{center}

\vspace{0.5cm}
\begin{center}
{\sc Martin~Beneke},$^{a}$ 
{\sc Mathias~Garny},$^{a}$
{\sc Sebastian Jaskiewicz},$^{a}$ \\ 
{\sc Robert~Szafron},$^{a,b}$
{\sc Leonardo Vernazza},$^{c,d,e}$ 
and {\sc Jian~Wang}$^{f}$\\[6mm]
{\it $^a$ Physik Department T31,\\
James-Franck-Stra\ss e~1, 
Technische Universit\"at M\"unchen,\\
D--85748 Garching, Germany\\[0.2cm]
$^b$ Theoretical Physics Department, CERN,\\ 
1211 Geneva 23, Switzerland\\[0.2cm]
$^c$ Institute for Theoretical Physics Amsterdam 
and \\
Delta Institute for Theoretical Physics, \\
University of Amsterdam, Science Park 904,\\
NL--1098 XH Amsterdam, The Netherlands\\[0.2cm]
$^d$ Nikhef, Science Park 105,\\ 
NL--1098 XG Amsterdam, 
The Netherlands\\[0.2cm]
$^e$Dipartimento di Fisica Teorica, 
Universit\`a di Torino, \\
and INFN, Sezione di Torino, Via P. Giuria 1, 
I-10125 Torino, Italy \\[0.2cm]
$^f$School of Physics, Shandong University,\\ 
Jinan, Shandong 250100, China\\[0.2cm]
}
\end{center}

\vspace{0.4cm}
\begin{abstract}
\vskip0.2cm\noindent
We sum the leading logarithms $\alpha_s^n \ln^{2 n-1}(1-z)$, 
$n=1,2,\ldots$ near the kinematic threshold $z=m_H^2/\hat{s}\to 1$ 
at next-to-leading power in the expansion in $(1-z)$ for 
Higgs production in gluon fusion. We highlight the new contributions compared to Drell-Yan production in quark-antiquark 
annihilation and show that the final result can be obtained 
to all orders by the substitution of the colour factor 
$C_F\to C_A$, confirming previous fixed-order results 
and conjectures. We also provide a numerical analysis 
of the next-to-leading power leading logarithms, which indicates 
that they are numerically relevant.
\end{abstract}
\end{titlepage}

\section{Introduction}
\label{sec:introduction}

The Higgs production cross section in gluon-gluon fusion is 
presently the most precisely computed observable in hadron-hadron 
collisions, as far as orders in perturbation theory 
(N$^3$LO \cite{Anastasiou:2015ema,Anastasiou:2016cez,Mistlberger:2018etf,Dulat:2018bfe}
 in the heavy-top approximation) 
and threshold resummation (N$^3$LL 
\cite{Moch:2005ky,Laenen:2005uz,Idilbi:2005ni,Idilbi:2006dg,Ahrens:2008nc,Bonvini:2014joa}) 
are concerned. In \cite{Beneke:2018gvs} we developed the 
framework for threshold resummation at next-to-leading 
power (NLP) using effective field theory, taking 
the first step 
beyond the leading-power resummation formalism in perturbative 
QCD \cite{Sterman:1986aj,Catani:1989ne}. We applied it to the 
summation of the leading logarithms (LL) in the classic Drell-Yan 
process $q\bar{q} \to \gamma^*+X$. Given the interest 
in Higgs production both phenomenologically and for 
applying new methods to high-order calculations, we discuss 
the similarities and differences of NLP threshold resummation 
for Higgs production in gluon fusion compared to the case 
of a virtual 
photon in this paper.\footnote{The result of \cite{Beneke:2018gvs} 
has been confirmed in \cite{Bahjat-Abbas:2019fqa} with the 
diagrammatic method \cite{Laenen:2010uz,Bonocore:2015esa,DelDuca:2017twk}
and extended to related processes, including Higgs production. See also 
\cite{Moult:2018jjd} for NLP resummation for event shapes.}
Although it is not the main focus of the work, we also 
provide a numerical analysis of the next-to-leading 
power leading logarithms, which indicates that they are numerically 
relevant.

The outline of this paper is as follows. 
In Sec.~\ref{sec:NLPfact} we derive the factorization formula 
for single Higgs production at NLP in soft-collinear 
effective theory (SCET), identify the sources of 
NLP LLs, and derive the hard, soft, and collinear functions 
needed for resummation with LL accuracy. The resummation of 
NLP LLs via renormalization-group 
equations is contained in Sec.~\ref{sec:LLresum}.  
In the same section we also expand the resummed result in 
$\alpha_s$, which provides both, a check of the resummed result 
by comparison with existing fixed-order expressions, and  
so far unknown logarithmic terms at higher order. Finally, 
in Sec.~\ref{sec:numerics} we perform a numerical study 
of the NLP-resummed cross section.

For the sake of avoiding repetition, we build on 
\cite{Beneke:2018gvs} for basic definitions related to threshold 
resummation and SCET. We also recommend consulting that paper for the 
general logic of deriving the NLP factorization and simplifications 
at the leading-logarithmic order before continuing with 
this one.

\section{Threshold factorization at NLP}
\label{sec:NLPfact}

We consider the process 
\be
A(p_A) + B(p_B) \to \mbox{H}(q)+X, 
\ee
where $A(p_A)$, $B(p_B)$ represent the colliding protons, and 
$X$ denotes an unobserved QCD final state. The cross section 
for this process can be written as 
\be \label{sigma-conv-renorm-explicit}
\sigma = \frac{\as^2}{576 \pi v^2} 
\sum_{a,b}  \int_0^1 dx_a \int_0^1 dx_b \,
f_{a/A}(x_a) f_{b/B}(x_b) \, \hat \sigma_{ab}(z),
\ee
where $f_{i/I}(x_i)$ represent parton distribution
functions and $v$ is the Higgs vacuum expectation value, 
$v^2 = 1/(\sqrt{2} G_F)$. $G_F$ is the Fermi 
constant, and $\alpha_s$ without scale argument 
refers to the strong coupling at 
the $\overline{\rm MS}$ scale $\mu$. 
In the following, we consider only the 
gluon-gluon initial state and drop the indices 
$a,b$. Higgs production in gluon fusion occurs through a top 
loop, which couples the gluons to the Higgs boson. In the 
heavy-top-quark mass limit  $m_t \gg m_H$, the Higgs boson 
couples to gluons via
\be\label{LHiggsEff}
{\cal L}_{\rm eff} = 
C_t\left(m_t,\,\mu\right)\, 
\frac{\as}{12\pi} \frac{H}{v} F^A_{\mu\nu}F^{\mu\nu A},
\ee
where 
\be \label{LHiggsEffC}
C_t\left(m_t,\,\mu\right) = 1
+\frac{\as}{4\pi} (5C_A - 3C_F)
 + \ord(\as^2)\,.
\ee

The partonic cross section $\hat{\sigma} \equiv \hat{\sigma}_{gg}$ 
is a function of the dimensionless variable $z=m_H^2/\hat{s}$, 
where $\hat{s} = x_a x_b \, s$ represents the partonic 
centre-of-mass energy squared, and  $x_a$, $x_b$ 
the momentum fractions of 
the gluons in the corresponding hadrons, defined through the 
parton momenta $p_a^\mu = x_a \sqrt{s} n_-^\mu/2$,  $p_b^\mu = 
x_b\sqrt{s} n_+^\mu/2$. Our aim is to sum the {\em leading 
logarithms} in the series $\sum_{n=1}^\infty \sum_{m=0}^{2 n-1} 
d_{nm} \ln^m(1-z)$ of {\em next-to-leading power} logarithms 
as was done for  $q\bar{q} \to \gamma^*+X$ \cite{Beneke:2018gvs}.
To this end, we start from the Higgs production cross section 
formula
\be\label{QCDxs}
\sigma = \frac{1}{2 s} 
\int \frac{d^3 \vec{q}}{(2\pi)^3 2E_q}
\sum_X |\langle H X| A B \rangle|^2
(2\pi)^4\delta^{(4)} (p_A + p_B - q  - p_X),
\ee
where the matrix element squared reads
\bea\label{QCD-matrix-element} \nn
\sum_X |\langle H X| AB \rangle|^2 &=&
\frac{\as^2(\mu)\, C^2_t(m_t,\mu)}{144\pi^2 v^2} \\
&&\hspace{-2.0cm}\times \, 
\sum_X \langle AB | \big[F^{A'}_{\rho\sigma}F^{\rho\sigma A'}
\big](0) |X \rangle 
\langle X|  \big[F^A_{\mu\nu}F^{\mu\nu A}\big](0) |AB \rangle. 
\eea
The QCD operator $F^A_{\mu\nu}F^{\mu\nu A}$ is then matched to 
SCET operators. At leading power (LP) this involves the single
operator
\begin{equation}
F^A_{\mu\nu}F^{\mu\nu A} (0) 
= \int dt\, d\bar{t}\, \widetilde{C}^{A0}(t,\bar{t}\,) 
\, J^{A0}(t,\bar{t}\,)\,,
\label{eq:A0current}
\end{equation}
where
\begin{eqnarray} 
&& \label{eq:A0operator} 
J^{A0}(t,\bar{t}\,) =  
2 g_{\mu\nu} \, n_- \partial {\cal A}^{\nu A}_{\,\bar{c} \perp}(\bar t n_-)
\, n_+ \partial {\cal A}^{\mu A}_{\,c \perp}(t n_+)\,, \\
&& \label{eq:A0Wilson}
C^{A0}(\np p, \nm \bar{p}\,) = \int dt\, d\bar{t}\, 
e^{-i (\np p) t -i ( \nm \bar{p})\bar{t}} \, 
\widetilde{C}^{A0}(t,\bar{t}\,) \,.
\end{eqnarray}
Here $\mathcal{A}_{c\perp}^\mu$ denotes the 
collinear-gauge-invariant transverse collinear gluon field of SCET. 
The operator generates a collinear and an anti-collinear field 
and is non-local along the light-rays in the respective 
directions. The derivatives correspond to the large momentum 
components of the fields. With the normalization 
(\ref{eq:A0operator}), the hard coefficient function 
is $\widetilde{C}^{A0}(t,\bar{t}\,) = \delta(t)\delta(\bar{t})+
\mathcal{O}(\alpha_s)$, or 
$C^{A0}(\np p, \nm \bar{p}) = 1+\mathcal{O}(\alpha_s)$
in momentum space. For LL resummation, the tree approximation 
will suffice. 

Beyond LP the hard gluon-gluon-Higgs vertex is matched 
to higher-power SCET operators. The basis of such operators 
in the position-space SCET formalism is discussed in \cite{Beneke:2017ztn,Beneke:2018rbh,Beneke:2019kgv}\footnote{For an operator basis at NLP in the alternative momentum-space SCET formulation see \cite{Marcantonini:2008qn,Kolodrubetz:2016uim,Feige:2017zci,Moult:2017rpl}.}. 
Following the same arguments as for $q\bar{q}\to \gamma^*$ 
\cite{Beneke:2018gvs,BBJVunpublished},  
to first subleading power in ($1-z$), i.e.~to order $\lambda^2$, 
the LLs arise only from the time-ordered product of the LP 
operator $J^{A0}$ with the $\mathcal{O}(\lambda^2)$ 
suppressed interactions from the SCET Lagrangian 
\cite{Beneke:2002ni}, 
\be\label{time-ordered-ps}
J_{A0,\,j}^{T2}(t,\bar{t}\,) = i \int d^4 z \, 
{\bf T}\Big[ J_{A0}(t,\bar{t}\,) \, {\cal L}^{(2)}_{j}(z) \Big], 
\ee
where the index $j$ labels terms in the power-suppressed
Lagrangian ${\cal L}^{(2)}(z)$. Thus, to obtain the NLP LL 
accurate amplitude, we simply add this operator to 
$J^{A0}(t, \bar{t}\,)$ in \eqref{eq:A0current}.

To proceed, one removes the soft-collinear interactions from the 
LP Lagrangian  by a field redefinition \cite{Bauer:2001yt}, in 
which (anti-) collinear gluon fields are multiplied 
by adjoint soft Wilson lines ${\cal Y}_+$ (${\cal Y}_-$), 
defined as 
\be \label{adj-soft-Wilson-lines}
{\cal Y}^{AB}_{\pm}(x) = {\bf P} \exp \bigg\{g_s \int _{-\infty}^0 ds 
\, f^{ABC}\, n_{\mp} A^C_s(x + s n_{\mp}) \bigg\}.
\ee
In terms of the decoupled collinear fields, which will be 
used below, the operator in \eqref{eq:A0operator} reads
\begin{equation}
J^{A0}(t,\bar{t}\,) = 
2 g_{\mu\nu} \,
{\cal Y}^{AC}_{-}(0) \, 
n_- \partial {\cal A}^{\nu C}_{\bar{c} \perp}(\bar t n_-) \,
{\cal Y}^{AD}_{+}(0)  \, 
n_+ \partial {\cal A}^{\mu D}_{c \perp}(t n_+) \,.
\label{eq:A0decoupled}
\end{equation}
The interaction Lagrangian is then also expressed in terms of 
the decoupled collinear field and the soft-gluon building 
block
\begin{eqnarray}
\mathcal{B}_{\pm}^{\mu}&=&Y_{\pm}^{\dagger}\left[iD_{s}^{\mu}Y_{\pm}\right]\,,
\end{eqnarray}
evaluated at the appropriately multipole-expanded position, 
where $Y_{\pm}(x)$ represent soft Wilson lines 
in the fundamental colour representation:
\be \label{soft-Wilson-lines}
Y_{\pm}(x) = {\bf P} \exp \bigg\{ i g_s 
\int _{-\infty}^0 ds \, n_{\mp} A_s(x + s n_{\mp}) \bigg\}.
\ee
An analysis of the interaction terms similar 
to \cite{Beneke:2018gvs}, now for the Yang-Mills SCET 
Lagrangian, shows that only the two terms 
\bea\label{Lym2} \nn
{\cal L}_{\rm 1YM}^{(2)} &=&
-\frac{1}{2g_s^2}\,{\rm tr} \,\Big( 
\big[ n_+ \partial \, {\cal A}^c_{\nu_{\perp}} \big] 
\Big[ n_-x\, i n_- \partial \, n_+{\cal B}^+, \, 
{\cal A}_c^{\nu_{\perp}} \Big] \Big),  \\ \nn
{\cal L}_{\rm 2YM}^{(2)} &=& 
- \frac{1}{2g^2_s}\,{\rm tr} \,\Big(
\big[ n_+ \partial \, {\cal A}^c_{\nu_{\perp}} \big]  \,
\Big[ x_{\perp}^{\rho} x_{\perp \omega} 
\big[ \partial^{\omega}, \, i n_- \partial \, {\cal B}^+_{\rho} \big],
\,  {\cal A}_c^{\nu_{\perp}} \Big] \Big)
\eea
are relevant for the leading logarithms. 

One of the main new features of the factorization theorem 
beyond LP is the appearance of collinear 
functions \cite{Beneke:2018gvs,BBJVunpublished}. They are defined 
as the perturbative matching coefficients of threshold-collinear 
fields with virtuality $m_H^2 (1-z) \gg \Lambda^2$ to 
PDF-collinear fields (the modes contained in the parton 
distribution function) with virtuality $\Lambda^2$, in the 
presence of soft fields with virtuality $m_H^2(1-z)^2$. By using the 
equation-of-motion identity 
\be
\label{eq:eom}
n_+ {\cal B}^+ = -2 \,\frac{i \partial_{\perp}^{\mu}}
{i n_- \partial}{\cal B}_{\perp \mu}^+ 
+ \mbox{two-parton terms},
\ee
we observe that $\mathcal{L}_{\rm 1 YM}^{(2)}$ 
and $\mathcal{L}_{\rm 2 YM}^{(2)}$ above can be written in terms
of the same soft building block. Hence there is only 
a single collinear function, defined through the 
relation\footnote{A similar definition applies to the 
anti-collinear gluon field with $\np$ and $\nm$ exchanged.}
\begin{eqnarray} 
\nn
i\int d^4z\,
\mathbf{T}\left[n_+ \partial {\cal A}^Y_{c \mu_\perp}
(t n_+) \Big(\mathcal{L}^{(2)}_{{\rm 1YM}}(z) +
\mathcal{L}^{(2)}_{{\rm 2YM}}(z)\Big)\right] && \\ 
&&\hspace{-8.0cm}
=\, 2\pi \int du \int \frac{d(n_+z)}{2} \,\widetilde 
J_{{\rm YM}\, \mu\rho}^{\,Y\!BC}\left(t,u;\frac{n_+z}{2}\right)
\, {\cal A}^{C \rho_\perp,\,\rm PDF}_{c}(u n_+) \, 
\frac{\partial^{\,\omega}_{\perp}}{in_-\partial}
\mathcal{B}^{+B}_{\omega_\perp}(z_-)\,.\quad
\label{collfn-1dot1}
\end{eqnarray}
Below we express the amplitude in terms of the Fourier 
transforms 
\be\label{gluonPDFcollinear}
{\cal \hat A}^{C,\,\rm PDF}_{\,c \alpha_\perp}(n_+p) 
= \int du \, e^{i (n_+ p)u} 
{\cal A}^{C,\,\rm PDF}_{\,c \alpha_\perp}(u n_+),
\ee
and 
\bea
J_{{\rm YM}\, \mu\rho}^{\,Y\!BC}\left(n_+ p,\np p';\omega\right) 
&=& 
\int dt \, e^{i (n_+ p) t} \, \int du \, e^{-i (n_+ p') u}
\nonumber\\
&&\times\,\int \frac{d(\np z)}{2} \, e^{i \omega (n_+ z)/2} \,
\widetilde{J}_{{\rm YM}\, \mu\rho}^{\,Y\!BC}
\left(t,u;\frac{n_+z}{2}\right).
\eea
With these definitions, the collinear 
function at the lowest order is easily calculated to be 
\be\label{collinear-res}
J_{{\rm YM}\,\mu\rho}^{Y\!BC}(n_+p, n_+ p';\omega) = 
-2i \,T_R\, f^{Y\!BC} \, g_{\perp \mu\rho} 
\bigg[ 2 - 2n_+ p'\frac{\partial}{\partial n_+ p} \bigg]
\delta(n_+ p - n_+ p')
\ee
with $T_R=1/2$. The Lagrangian insertion  
$\mathcal{L}_{\rm 1 YM}^{(2)}$ contributes $1 - 
2n_+ p'\frac{\partial}{\partial n_+ p}$ to this result, 
while the remaining $1$ is due 
to $\mathcal{L}_{\rm 2 YM}^{(2)}$. For the LL resummation, 
the lowest order, tree-level expression for the collinear 
function suffices.

At this point, we can put together previous expressions 
to write the NLP contribution to the matrix element 
$\langle X|\big[F^A_{\mu\nu}F^{\mu\nu A}\big](0) |AB \rangle$, 
which appears in (\ref{QCD-matrix-element}), in the factorized form 
\bea\label{NLPme-b} \nn
\langle X | \left[ F^A_{\mu\nu}F^{\mu\nu A} \right](0)\,  
| A(p_A) B(p_B) \rangle_{\rm NLP} &=& 
-2i \, \int \frac{dn_+ p}{2\pi} \frac{dn_- \bar p}{2\pi} 
\, g^{\mu\nu} \, C^{A0}(n_+p,n_- \bar p)  
\\ 
&&\hspace{-5.0cm}\times \, \int dn_- p_b \, 
\delta(n_-\bar p - n_- p_b) \,n_- p_b \,
\langle X_{\bar c_{\rm PDF}} | 
{\cal \hat A}^{X,\,\rm PDF}_{\,\bar{c} \nu_\perp}(n_- p_b) | B(p_B) 
\rangle 
\nn \\ 
&&\hspace{-5cm} \times \, 
\int dn_+ p_a \, \langle X_{c_{\rm PDF}} | 
{\cal \hat A}^{C \rho_\perp,\,\rm PDF}_{c}(n_+ p_a)|A(p_A) \rangle \,
\int \frac{d\omega}{4\pi} \,
J_{{\rm YM}\,\mu\rho}^{\,Y\!BC}(n_+p, n_+ p_a;\omega)  
\\ 
&&\hspace{-5.0cm} \times \, 
\int d(n_+ z) \,e^{-i \omega (n_+ z)/2} \,
\langle X_s | {\bf T} \bigg[ 
{\cal Y}^{\,AX}_{-}(0) {\cal Y}^{\,AY}_{+}(0) 
\, \frac{\partial^{\,\omega}_{\perp}}{in_-\partial}
\mathcal{B}^{+B}_{\omega_\perp}(z_-) \bigg] | 0 \rangle 
+\bar c\mbox{-term}\,,\quad
\nn\eea
where we used that the final state $\langle X|$ contains 
threshold-soft and c-PDF states, 
$\langle X_s| \otimes \langle X_{c,{\rm PDF}}| 
\otimes \langle X_{\bar c,{\rm PDF}}|$. 
Integrating by parts the derivative in the 
collinear function \eqref{collinear-res}, it  
acts on the rest of the matrix element. In \eqref{NLPme-b}, 
the only term which depends on $n_+ p$ is the short-distance
coefficient $C^{A0}(n_+p,n_- \bar p)$. With the normalization 
adopted in \eqref{eq:A0current} one has $C^{A0}(n_+p,n_- \bar p) 
= 1 + {\cal O}(\as)$, and since we will need only the tree-level 
expression, the derivative term in 
\eqref{collinear-res} does not contribute. We therefore 
extract the momentum, colour, and Lorentz structure of 
$J_{\mu\rho}^{Y\!BC}(n_+p, n_+ p';\omega)$ and 
substitute
\be\label{collinear-res-b}
J_{{\rm YM}\,\mu\rho}^{Y\!BC}(n_+p, n_+ p_a;\omega) \;\to\;  
i f^{Y\!BC} \, g_{\perp \mu\rho} \, J_{\rm YM}(n_+ p_a;\omega)
\, \delta(n_+ p - n_+ p_a),
\ee
where, at the lowest order
\be\label{coll0}
J_{\rm YM}^{(0)}(n_+ p_a;\omega) = - 4 T_R = -2.
\ee
Inserting \eqref{collinear-res-b} into 
\eqref{NLPme-b} we get 
\bea\label{NLPme-c} \nn
\langle X | \left[ F^A_{\mu\nu}F^{\mu\nu A} \right](0)\,  
| A(p_A) B(p_B) \rangle_{\rm NLP} 
&=& 
 2  \int \frac{dn_+ p_a}{2\pi} \frac{dn_- p_b}{2\pi} 
\, g^{\,\mu\nu} \, C^{A0}(n_+p_a,n_- p_b)  
\\ \nn
&&\hspace{-5.0cm}\times  \, 
n_- p_b \,\langle X_{\bar c_{\rm PDF}} | 
{\cal \hat A}^{X,\,\rm PDF}_{\,\bar{c} \nu_\perp}(n_- p_b) | B(p_B) 
\rangle
\, \langle X_{c_{\rm PDF}} | 
{\cal \hat A}^{C,\,\rm PDF}_{\,c \mu_\perp}(n_+ p_a) | A(p_A) 
\rangle \\ 
&&\hspace{-5.0cm} \times \, 
\int \frac{d\omega}{4\pi} 
J_{\,\rm YM}(n_+ p_a;\omega) 
\int d(n_+ z) \,e^{-i \omega (n_+ z)/2} \\ \nn
&&\hspace{-5cm} \times \, 
\langle X_s | {\bf T} \bigg[
{\cal Y}^{AX}_{-}(0) {\cal Y}^{AY}_{+}(0) 
\, f^{Y\!BC} \,\frac{\partial^{\,\omega}_{\perp}}{in_-\partial}
\mathcal{B}^{+B}_{\omega_\perp}(z_-) \bigg] | 0 \rangle +\bar c\mbox{-term}.
\eea
Upon comparison with Eq.~(3.17) of \cite{Beneke:2018gvs},
we see that the matrix element for Higgs production at 
NLP is very similar to the one obtained for the production 
of a virtual photon, 
with obvious differences in the colour structure, 
related to the gluonic instead of the quark-antiquark 
initial state.
At leading logarithmic accuracy one can further 
set $C^{A0}(\np p_a,\nm p_b)\to 1$ at the 
hard scale, and $J_{\rm YM}(n_+ p_a;\omega) \to - 2$ 
at the collinear scale, which further simplifies 
(\ref{NLPme-c}).

In the next step, we square the matrix element to obtain the 
factorized form of (\ref{QCDxs}), (\ref{QCD-matrix-element}). 
Summation of the PDF-(anti-)collinear final state introduces 
the gluon parton distribution
\bea\label{PDF-gluon} \nn
\big\langle A(p_A) \big| {\cal A}^{A',\,\rm PDF}_{\,c \rho'_\perp}
(x+ u' n_+) 
{\cal A}^{A,\,\rm PDF}_{\,c \rho_\perp}(u n_+) 
\big| A(p_A)\big\rangle
&=& -\frac{g_{\perp\rho\rho'}}{2}
\frac{\delta^{AA'}}{N^2_c-1} \\
&&\hspace{-3.0cm}\times  \, \int_{0}^{1} 
\frac{d x_a}{x_a} \, f_{g/A}(x_a) \, e^{i x_a (x + u' n_+ - u n_+)\cdot p_A}\,,
\qquad
\eea
while the sum over the soft radiation yields the NLP soft 
function $S_{\rm YM}(\Omega,\omega)$ defined below.

Performing the integrations over $\np p_a,\nm p_b$ and 
stripping off the convolution with the gluon distribution 
functions, we obtain the partonic cross section 
\eqref{sigma-conv-renorm-explicit} up to NLP in 
the threshold expansion (including the LP term) in the 
form
\begin{eqnarray}
\hat{\sigma}(z) &=&  
\frac{8 C_t^2(m_t)}{N_c^2-1}\,\hat s H(\hat{s})
\int\frac{d^3\vec{q}}{(2\pi)^3 \,2\sqrt{m_H^2+\vec{q\,}^{\,2}}}\,
\frac{1}{2\pi} \int d^4x\,
e^{i(x_a p_A+x_b p_B-q)\cdot x}
\nonumber \\
&&\times\left\{
\widetilde{S}_0(x) 
- \frac{2}{\sqrt{\hat s}} 
\int d\omega \, J_{\rm YM}(x_a \np p_A;\omega)\,
\widetilde{S}_{\rm YM}(x,\omega) + \,\mbox{$\bar{c}$-term}\,
\right\}\,.\qquad
\label{eq:NLPfactgg}
\end{eqnarray}
Here 
\begin{equation}
H(\hat{s},\mu_h) =  |C^{A0}(-\hat{s})|^2\,,
\end{equation}
represents the hard function, which is the same 
for the LP and NLP term. $\widetilde{S}_0(x)$ 
denotes the LP position-space soft function of adjoint Wilson 
lines for the gluon-gluon initial state
generalized to $x^0\to x^\mu = (x^0,\vec{x}\,)$ in 
the position argument of the Wilson lines. Its Fourier transform 
with respect to $x^0$ will be denoted by $S_0(\Omega,\vec{x})$, 
such that $S_{\rm H}(\Omega) = S_0(\Omega,\vec{0}\,)$.
Furthermore, $\widetilde{S}_{\rm YM}(x,\omega)$ 
represents the NLP soft function, defined as the 
Fourier transform
\begin{equation}
\widetilde{S}_{\rm YM}(x,\omega) = 
\int\frac{d(\np z)}{4\pi}\,
e^{- i\omega (\np z)/2} \,
\frac{1}{N_c^2-1}\,
\langle 0 |\widetilde{\mathcal{S}}_{\rm YM}(x,z_-)|0\rangle\,,
\label{eq:s2xitraced}
\end{equation}
of the vacuum matrix element of the operator 
\begin{equation}
\widetilde{\mathcal{S}}_{\rm YM}\left(x,z_-\right)
= {\bf \bar T} \Big[ {\cal Y}^{\,A'C}_{+}(x)
 {\cal Y}^{\,A'X}_{-}(x) \Big]
{\bf T} \Big[ {\cal Y}^{\,AX}_{-}(0)
 {\cal Y}^{\,AY}_{+}(0) \,  f^{Y\!BC} \, 
\frac{\partial^{\,\sigma}_{\perp}}{i n_- \partial} 
{\cal B}_{\sigma_\perp}^{+B}(z_-) \Big].
\label{eq:softdef}
\end{equation}
We will denote the Fourier transform of $\widetilde{S}_{\rm YM}(x,
\omega)_{|\vec{x}=0}$ with respect to $x^0$ 
by $S_{\rm YM}(\Omega,\omega)$. The factor of two multiplying 
$J_{\rm YM} \otimes \widetilde S_{YM}$ in \eqref{eq:NLPfactgg} 
arises from the two identical NLP terms in the 
square of the amplitude. 

As discussed in \cite{Beneke:2018gvs}, a number of ``kinematic'' 
power corrections arise from expanding the first line of 
(\ref{eq:NLPfactgg}) and the generalized LP soft function
$\widetilde{S}_0(x)$. We shall also consider the partonic 
cross section rescaled by a factor of $1/z$, 
\begin{equation}
\Delta(z) =\frac{\hat{\sigma}(z)}{z} , 
\label{eq:delta}
\end{equation}
 as is conventionally done. The 
derivation of the kinematic correction is almost identical 
to the $q\bar{q}\to\gamma^*$ case, and we refer to 
\cite{Beneke:2018gvs} for further details. A difference 
arises from the factor $\hat{s} = m_H^2/z$ in (\ref{eq:NLPfactgg}), 
which is absent in $\gamma^*$ production, and which originated 
from the derivatives in the Higgs production operator 
(\ref{eq:A0operator}). Together with the $1/z$ factor from 
(\ref{eq:delta}) this implies that the kinematic correction 
denoted by $S_{K3}(\Omega)$ in \cite{Beneke:2018gvs} 
is twice as large, and hence the sum of all kinematic 
corrections does no longer cancel for the quantity 
$\Delta(z)$ at LL accuracy. Instead we obtain 
\begin{eqnarray} \nn
\Delta(z) &=&\frac{8 C_t^2(m_t)}{N_c^2-1}
\,m_H \,H(m_H^2)
\bigg\{S_{\rm H}(m_H(1-z)) 
+ \frac{1}{m_H} S_{K}(m_H(1-z)) \\
&&\hspace{1.5cm}-\,  \frac{2}{m_H} 
\int d\omega \, J_{\rm YM}(x_a \np p_A;\omega)\,
S_{\rm YM}(m_H(1-z),\omega) + \,\mbox{$\bar{c}$-term}\,
\bigg\}\,,\qquad 
\label{eq:NLPfactggB}
\end{eqnarray}
with 
\begin{equation}
S_{K}(\Omega) = 
\frac{\alpha_{s}C_{A}}{2\pi}\left(
-8 \ln\frac{\mu}{\Omega} + 4\right)\theta(\Omega)
+ \mathcal{O}(\alpha_s^2)
\,.
\end{equation}
Hence, in the case of Higgs production 
the kinematic corrections \emph{do} produce 
NLP LLs in $\Delta(z)$. 

\section{Resummation}
\label{sec:LLresum}

The resummation of NLP logarithms is performed using renormalization 
group equations (RGEs) to evolve the scale-dependent functions in 
the factorization formula \eqref{eq:NLPfactggB} to a common scale, 
for which we adopt the collinear scale $\mu_c\sim m_H\sqrt{1-z}$. 
One difference with respect to the classic DY process is due to the 
effective $ggH$ vertex, which introduces the additional 
short-distance coefficient $C_t(m_t)$, which multiplies 
both, the  LP and NLP term. The value of $C_t$ at a generic scale 
$\mu$ is (see, for example  \cite{Ahrens:2008nc}) 
\be\label{eq:CtRGE}
C_t(m_t,\mu) = 
\frac{\beta\big(\as(\mu)\big)}{\as^2(\mu)}
\frac{\as^2(\mu_t)}{\beta\big(\as(\mu_t)\big)}\,
C_t(m_t,\mu_t),
\ee
where $C_t(m_t,\mu_t)$ gives the initial condition
at the scale $\mu_t \sim m_t$, and 
\begin{equation}
\beta(\alpha_s) = \frac{d}{d\ln\mu}\alpha_s = 
-2\,\frac{\beta_0\alpha_s^2}{4\pi} 
+\mathcal{O}(\alpha_s^3),
\qquad
\beta_0=\frac{11}{3} N_c-\frac{2}{3}n_f\,.
\end{equation}
 Next, we need to consider
the evolution of the hard function $H(m_H^2)$ from the hard scale 
$\mu_h\sim m_H$ to the collinear scale. This is identical 
to the $q\bar{q}$ case \cite{Beneke:2018gvs}, up to the 
colour-factor substitution $C_F\to C_A$. To LL accuracy one 
has 
\begin{equation}
H(m_H^2,\mu) = \exp\left[4 S^{\rm LL}(\mu_h,\mu)\right] 
H(m_H^2,\mu_h) \,,
\label{eq:hardRGE}
\end{equation}
where 
\be
S^{\rm LL}(\nu,\mu) =\frac{C_A}{\beta_0^2} 
\frac{4\pi}{\alpha_s(\nu)}
\left(1-\frac{\alpha_s(\nu)}{\alpha_s(\mu)}
+\ln \frac{\alpha_s(\nu)}{\alpha_s(\mu)}\right)\,.
\label{eq:SLL}
\ee

Regarding the evolution of the NLP soft function 
(\ref{eq:s2xitraced}), (\ref{eq:softdef}) from the soft 
scale $\mu_s\sim m_H (1-z)$ to the collinear scale, we first 
note that this soft function has exactly the same form 
as the one that appeared for $q\bar{q}\to\gamma^*$ in 
\cite{Beneke:2018gvs} with the only difference that the 
Wilson lines and colour operators are now in the adjoint 
rather than in the fundamental representation. The structure 
of the RGE relevant to LL 
resummation, as well as the $\mathcal{O}(\alpha_s)$ 
result for the soft function follows from substituting 
$C_F\to C_A$ in the corresponding expressions in 
\cite{Beneke:2018gvs}. In particular, the LLs are generated 
from mixing between  $\widetilde{S}_{\rm YM}(x,\omega)$ and 
\be\label{eq:x0softfn}
S^{\rm ad}_{x_0}(\Omega) = \int\frac{dx^0}{4\pi}\,e^{ix^0\Omega/2}
\,\frac{-2i}{x^0-i\varepsilon}\, \frac{1}{N_c^2-1} \,
\langle 0 | {\bf \bar T} \Big[ {\cal Y}^{A'Y}_{+}(x^0) {\cal Y}^{A'X}_{-}(x^0) \Big] 
{\bf T} \Big[ {\cal Y}^{AX}_{-}(0) {\cal Y}^{AY}_{+}(0) \Big] | 
0 \rangle,
\ee
which is the adjoint-representation equivalent to 
$S_{x_0}(\Omega)$ defined in \cite{Beneke:2018gvs}.
We refer to \cite{Beneke:2018gvs} for the renormalization 
of these soft functions, which implies the RGE system
\begin{equation}
\frac{d}{d\ln\mu}\left(\begin{array}{c}
S_{\rm YM}(\Omega,\omega) \\
S_{x_{0}}^{\rm ad}(\Omega)
\end{array}\right)=\frac{\alpha_{s}}{\pi}
\left(\begin{array}{cc}
4C_{A}\ln\displaystyle\frac{\mu}{\mu_s}
& \quad \!\!C_{A}\delta(\omega) \\
0 & \quad \!4C_{A}\ln\displaystyle\frac{\mu}{\mu_s}
\end{array}\right)\left(\begin{array}{c}
S_{\rm YM}(\Omega,\omega)\\
S^{\rm ad}_{x^{0}}\left(\Omega\right)
\end{array}\right)\,,
\label{eq:ADMcopy}
\end{equation}
where $\mu_s$ denotes an arbitrary soft scale of 
order $m_H (1-z)$. In case of Higgs production the evolution of 
the kinematic soft function $S_{\rm K}$ must also be 
considered, and we find 
\begin{equation}
\frac{d}{d\ln\mu}\left(\begin{array}{c}
S_{\rm K}(\Omega) \\
S^{\rm ad}_{x_{0}}\left(\Omega\right)
\end{array}\right)=\frac{\alpha_{s}}{\pi}\left(\begin{array}{cc}
4C_{A}\ln\displaystyle\frac{\mu}{\mu_s} & 
\quad \!\! -4C_{A} \\
0 & \quad \!\! 4C_{A}\ln\displaystyle\frac{\mu}{\mu_s}
\end{array}\right)\left(\begin{array}{c}
S_{\rm K}(\Omega) \\
S^{\rm ad}_{x^{0}}\left(\Omega\right)
\end{array}\right)\,.
\label{eq:ADMcopy-kin}
\end{equation}
Following appendix A of \cite{Beneke:2018gvs} 
we obtain the LL solution
\begin{eqnarray} \nn
S_{\rm K}^{\rm LL}(\Omega,\mu) &=&
\frac{8 C_A}{\beta_0}
\ln\frac{\alpha_s(\mu)}{\alpha_s(\mu_s)}\,
\mbox{exp}\left[-4 S^{\rm LL}(\mu_s,\mu)\right]
\,\theta(\Omega)\,, \\
S_{\rm YM}^{\rm LL}(\Omega,\omega,\mu) &=&-
\frac{2 C_A}{\beta_0}
\ln\frac{\alpha_s(\mu)}{\alpha_s(\mu_s)}\,
\mbox{exp}\left[-4 S^{\rm LL}(\mu_s,\mu)\right]
\,\theta(\Omega)\delta(\omega)\,.
\label{eq:LLsolNLP}
\end{eqnarray}

Since the collinear function at the collinear scale does not 
contain large logarithms, we can insert the tree-level 
expression \eqref{coll0} into \eqref{eq:NLPfactggB} to 
obtain\footnote{The following equation is written under the 
assumption that we do not distinguish the scale of the effective 
Higgs-gluon coupling and the SCET factorization scale. The SCET 
anomalous dimension that governs the evolution of $H(m_H^2,\mu)$ then 
inherits a contribution from the anomalous dimension of 
the $HFF\,$ operator, which compensates the evolution of 
$C_t^2(m_t,\mu)$ below the hard scale. For the conceptually 
cleaner treatment of distinguishing the two scales, see 
the discussion of tensor quark currents in \cite{Bell:2010mg}. 
The final result (\ref{eq:NLPsummedfinal}) is the same 
in both cases.}
\begin{eqnarray}
\nn
\Delta(z,\mu_c) &=& \frac{\as^2(\mu_c)}{\as^2(\mu)} \,
C^2_t(m_t,\mu_c) \, H(m_H^2,\mu_c) \,\bigg\{
m_H\, S_{\rm H}\big(m_H (1-z),\mu_c\big) + S_{\rm K}^{\rm LL}(\Omega,\mu_c) \\
&&\hspace{0.0cm} 
+ 8 \int d\omega \,S^{\rm LL}_{\rm YM}\big(m_H(1-z),\omega,\mu_c\big) \bigg\}
\label{eq:NLPfactevolved2}
\end{eqnarray}
in terms of evolved hard and soft functions. 
The factor $\as^2(\mu_c)/\as^2(\mu)$ in front is 
included to compensate for the fact that the hadronic 
cross section \eqref{sigma-conv-renorm-explicit}
contains the factor $\as^2(\mu)$ not included 
into $\Delta(z,\mu_c)$. 
Also, the $\bar c$-term in \eqref{eq:NLPfactggB} gives an 
identical NLP contribution to the collinear one, thus we 
take it into account by multiplying the second line 
of (\ref{eq:NLPfactevolved2}) by 
two. Inserting now the resummed soft functions 
\eqref{eq:LLsolNLP} into \eqref{eq:NLPfactevolved2},
and using $H(m_H^2,\mu_h) = 1+\mathcal{O}(\alpha_s)$,
we get
\begin{eqnarray} \nn
\Delta^{\rm LL}(z,\mu_c) &=&
\Delta^{\rm LL}_{\rm LP}(z,\mu_c)- 
\frac{\as^2(\mu_c)}{\as^2(\mu)} \,
\bigg[\frac{\beta\big(\as(\mu_c)\big)}{\as^2(\mu_c)}
\frac{\as^2(\mu_t)}{\beta\big(\as(\mu_t)\big)}\bigg]^2 
\,C^2_t(m_t,\mu_t) \\
&&\hspace{0.0cm}  \times \,
\mbox{exp}\left[4 S^{\rm LL}(\mu_h,\mu_c)-4 S^{\rm LL}(\mu_s,\mu_c)\right] \frac{8C_A}{\beta_0}
\ln\frac{\alpha_s(\mu_c)}{\alpha_s(\mu_s)}\,\theta(1-z)\,.
\quad
\label{eq:NLPsummedfinal}
\end{eqnarray}
It is remarkable that the kinematic and NLP soft
function contribution sum up to give a NLP resummation 
formula which is identical to the $q\bar{q}$ induced DY process, 
with the color factor $C_F$ replaced by $C_A$. Within the 
approach presented here the main differences between the 
$q\bar{q}$ and $gg$ channel appear in a) a factor of two difference 
in the collinear function due to the contribution of 
two Lagrangian terms (\ref{Lym2}) and b) the existence of 
a kinematic correction. Both differences are related to the 
derivatives in the production operator (\ref{eq:A0operator}), 
but cancel to produce the above result.

In \eqref{eq:NLPsummedfinal} the term $\Delta^{\rm LL}_{\rm LP}(z)$ 
represents the LL-resummed LP partonic cross section, in 
the present formalism given in \cite{Ahrens:2008nc}. We can set 
$\mu_h=m_H$, $\mu_s=m_H(1-z)$ and $\mu_c=m_H\sqrt{1-z}$, since the 
precise choice is irrelevant for the LLs. However, 
Eq.~\eqref{eq:NLPsummedfinal} is not yet of the most general 
form, since it implies that the factorization scale $\mu$ is  
set to $\mu_c=m_H\sqrt{1-z}$ in the parton distributions. We
translate the result to arbitrary $\mu$ by using the scale 
invariance of the hadronic cross section, 
as discussed in \cite{Beneke:2018gvs}. The 
result is that the functional form of the resummed 
partonic cross section at a generic scale $\mu$ is identical 
to the functional form at the scale $\mu_c$: 
\bea\label{eq:NLPsummedfinalgeneral} \nn
\Delta^{\rm LL}_{\rm NLP}(z,\mu) &=&
\bigg[\frac{\beta\big(\as(\mu)\big)}{\as^2(\mu)}
\frac{\as^2(\mu_t)}{\beta\big(\as(\mu_t)\big)}\bigg]^2 
\,C^2_t(m_t,\mu_t)  \\
&&\hspace{0.0cm} \times \, 
\mbox{exp}\left[4 S^{\rm LL}(\mu_h,\mu)-4 S^{\rm LL}(\mu_s,\mu)
\right]\,
\frac{-8C_A}{\beta_0} \ln\frac{\alpha_s(\mu)}{\alpha_s(\mu_s)}\,
\theta(1-z)\,,\quad
\eea
which implies that the collinear function cannot contain 
LLs when evaluated at a scale $\mu$ different 
from $\mu_c$. The scale of the parton luminosity that multiplies 
\eqref{eq:NLPsummedfinalgeneral} is now manifestly independent of 
$z$, and the logarithms of $(1-z)$ are generated by setting 
$\mu_s\sim m_H(1-z)$. 

Eq.~\eqref{eq:NLPsummedfinalgeneral} expresses the 
resummation of NLP LLs for Higgs production in gluon 
fusion, which constitutes our main result. We can expand 
it to fixed order in perturbation theory, to obtain the logarithms 
explicitly and to compare with existing results. Given that 
$\big[\beta\big(\as(\mu)\big)/\as^2(\mu)
\, \as^2(\mu_t)/\beta\big(\as(\mu_t)\big)\big]^2$ 
and $C^2_t(m_t,\mu_t)$ both equal unity at $\mathcal{O}(\alpha_s^0)$ 
and do not contain LLs in higher orders, we 
can drop these two factors altogether. The expansion of 
\eqref{eq:NLPsummedfinalgeneral} to fixed order is 
therefore the same as in $q\bar{q}\to \gamma^*$, with $C_F \to C_A$.
For arbitrary $\mu$, with $\mu_h=m_H$ and $\mu_s=m_H(1-z)$ 
we have 
\begin{eqnarray}
\Delta^{\rm LL}_{\rm NLP}(z,\mu) & =&
 {} - \theta(1-z)\,\bigg\{4C_A\frac{\alpha_s}{\pi}\Big[\ln(1-z)-
L_\mu \,\Big]\nn\\
&& \hspace*{-1.5cm} + \,8C_A^2\left(\frac{\alpha_s}{\pi}\right)^2
\Big[\ln^3(1-z)-3 L_\mu \ln^2(1-z)+2 L_\mu^2 \ln(1-z)\,\Big]
\nn\\
&&  \hspace*{-1.5cm}+ \,8C_A^3\left(\frac{\alpha_s}{\pi}\right)^3
\Big[\ln^5(1-z)-5 L_\mu \ln^4(1-z) +8 L_\mu^2 \ln^3(1-z) 
-4 L_\mu^3 \ln^2(1-z)\,\Big]
\nn\\
&&  \hspace*{-1.5cm}+ \,\frac{16}{3}C_A^4\left(\frac{\alpha_s}{\pi}\right)^4
\Big[\ln^7(1-z)-7 L_\mu\ln^6(1-z) +18 L_\mu^2 \ln^5(1-z) 
-20 L_\mu^3 \ln^4(1-z)\nn\\
&& + \, 8L_\mu^4 \ln^3(1-z)\,\Big]
\nn\\
&&  \hspace*{-1.5cm}+ \,
\frac{8}{3}C_A^5\left(\frac{\alpha_s}{\pi}\right)^5
\Big[\ln^9(1-z)-9 L_\mu\ln^8(1-z) +32 L_\mu^2 \ln^7(1-z) 
 -56L_\mu^3 \ln^6(1-z)\nn\\
&&
+\,48 L_\mu^4 \ln^5(1-z)-16L_\mu^5 \ln^4(1-z)
\,\Big]\bigg\} + \,{\cal O}(\alpha_s^6\times (\log)^{11})\,,
\label{eq:fullexpanded}
\end{eqnarray}
where we have defined $L_\mu=\ln(\mu/m_H)$, and $(\log)^{11}$ 
stands for some combination of the two logarithms to the 11th 
power.

The N$^3$LO term has been given by means of an exact calculation 
in \cite{Anastasiou:2014lda}, and also in \cite{deFlorian:2014vta}, 
based on the ``physical evolution kernels'' method, see in 
particular Eq.~(2.12) in \cite{Anastasiou:2014lda} and (B.2)
in \cite{deFlorian:2014vta}. Our result at this order agrees 
with these references. Furthermore, Eq.~(B.3) of 
\cite{deFlorian:2014vta} provides the result at N$^4$LO, 
with which we also agree.  The N$^5$LO term is a new result and the 
expansion to any order can be obtained without effort from 
\eqref{eq:NLPsummedfinalgeneral}. It proves the 
conjecture \cite{Kramer:1996iq,Kidonakis:2007ww} that 
the leading logarithm can be simply obtained from 
including the NLP term in the Altarelli-Parisi 
splitting kernels in the standard LP resummation 
formalism.

\section{Numerical analysis of NLP 
LL resummation}
\label{sec:numerics}

In this section, we provide a numerical exploration
of the NLP resummed Higgs production cross section in the 
large top mass approximation, via
gluon-gluon fusion at the LHC with $\sqrt{s}=13\,$TeV, and
using $m_H=125$\,GeV. The cross section is given by
\be \label{sigma-H}
\sigma =  \frac{\as^2(\mu) \, m_H^2}{576 \, \pi \, v^2\, s}
\int_{\tau}^1 \frac{dz}{z} \,
{\cal L} \bigg(\frac{\tau}{z},\mu \bigg)\,
\Delta(z,\mu),
\ee
where $\Delta(z,\mu)$ is related to the normalized partonic
cross section as defined in~(\ref{eq:delta}), and
the luminosity function involving the parton distribution functions
is given by
\be\label{luminosity}
{\cal L}(y,\mu) = \int_y^1\frac{dx}{x} 
f_{g/A}(x,\mu) f_{g/B}\left(\frac{y}{x},\mu\right).
\ee
We use the PDF sets PDF4LHC15nnlo30 
\cite{Buckley:2014ana,Butterworth:2015oua,Dulat:2015mca,Harland-Lang:2014zoa,Ball:2014uwa}.
For comparison, we also consider the resummed leading power cross section at NNLL accuracy, 
as provided in (30), (31) of \cite{Ahrens:2008nc}  (note that we strictly include only 
leading power contributions to the latter). The LP result involves 
hard and soft functions at one loop, the anomalous dimension 
$\Gamma_{\rm cusp}$ at three loops, and all other anomalous 
dimensions at two loops, see e.g. table 1 of \cite{Becher:2007ty}. 

The resummation formula at NLP depends on the 
scales $\mu_t$, $\mu_h$, $\mu_c$ and $\mu_s$, as well as on the factorization scale $\mu$. 
We choose $\mu_t =  173.1\,$GeV 
and $\mu_h = \mu = m_H$. In what follows, we consider also the
choice $\mu_h^2 = - m_H^2-i\epsilon$ (we omit $-i\epsilon$ below), which includes in the resummation factors of $\pi^2$ associated to
logarithmic contributions evaluated with time-like momentum transfer~\cite{Ahrens:2008nc}.
As discussed above, the NLP cross section does
not depend explicitly on the collinear scale $\mu_c$ at LL accuracy.
In section \ref{sec:LLresum} we used the parametric choice $\mu_s= m_H(1-z)$
for the soft scale 
to obtain analytic fixed-order results. We find that, as is known at LP, this choice
is not admissible to evaluate the resummed result. Instead, we use a 
dynamical soft scale given by~\cite{Sterman:2013nya}
\be \label{eq:softscale}
\mu^{\rm dyn}_s=\frac{Q}{\bar{s}_1(\tau)}, \quad \bar{s}_1(\tau) \equiv
-e^{\gamma_E}\frac{d\ln \mathcal{L} (y,\mu)}{d\ln y}\bigg|_{y=\tau}.
\ee
When applied to Higgs production at LP, this prescription gives a similar soft scale as the one used 
in \cite{Ahrens:2008nc}. Furthermore, it can be extended to NLP in a straightforward way.
With the Higgs mass and centre of mass energy given above 
we find $\mu_s^{\rm dyn} \simeq 38$ GeV. 
Alternatively, the resummation could be performed in Mellin space~\cite{Catani:1996yz,Moch:2005id}, 
which may be viewed as an implicit way to set an effective soft scale. 
Here we prefer to take advantage of keeping the soft scale independent from the
other scales in the problem, and consider the effect of varying $\mu_s$  below.

In table \ref{tablefix}, we consider the
perturbative expansion of the resummed cross section in order to investigate
whether our choice for the soft scale is suitable. To obtain the numerical values in table \ref{tablefix},
we evaluate both the LP and NLP result at LL accuracy, set the coefficient 
$C^2_t(m_t,\mu_t)$ together with its evolution factor (first line in~\eqref{eq:NLPsummedfinalgeneral}) to unity, and 
consider the running coupling constant at one loop. 
When setting the soft scale to its parametric value 
$\mu_s= m_H(1-z)$ 
the series is numerically divergent both at LP and NLP, 
as expected  (second and fifth column in table \ref{tablefix}). 
For $\mu_s = \mu^{\rm dyn}_s$, the higher order terms 
become suppressed (third and sixth column in table \ref{tablefix}), 
indicating that the expansion of the series is perturbatively
convergent for both the LP and NLP result, as required.
We also show the expanded result obtained for $\mu_h^2 = - m_H^2$
(fourth and seventh column in table 
\ref{tablefix}). As expected, the different choice 
of the hard scale does not alter the convergence of the LL
approximation. 

\begin{table}[t]
\begin{center}
\begin{tabular}{|c|r|r|r|r|r|r|}
\hline
(pb)    & \multicolumn{3}{c|}{$\sigma^{\rm LL}_{\rm LP}$ }&  
	      \multicolumn{3}{c|}{$\sigma^{\rm LL}_{\rm NLP}$ } \\ \hline
           & {\footnotesize $\mu_s = m_H(1-z)$} & {\footnotesize $\mu_s = \mu_s^{\rm dyn}$} & {\footnotesize $\mu_s = \mu_s^{\rm dyn}$}
           & {\footnotesize $\mu_s = m_H(1-z)$} & {\footnotesize $\mu_s = \mu_s^{\rm dyn}$} & {\footnotesize $\mu_s = \mu_s^{\rm dyn}$} \\
           & {\footnotesize $\mu_h^2 = m_H^2$} & {\footnotesize $\mu_h^2 = m_H^2$}  & {\footnotesize $\mu_h^2 = -m_H^2$} 
           & {\footnotesize $\mu_h^2 = m_H^2$} & {\footnotesize $\mu_h^2 = m_H^2$}  & {\footnotesize $\mu_h^2 = -m_H^2$} \\  \hline
$\ord\left(\as^0\right)$    &    12.94  &   12.94  &  12.94   &      --     &    --      &   --        \\
$\ord\left(\as\right)$        &      4.70  &     1.95  &    8.82   &       4.35   &     3.57    &   3.57   \\
$\ord\left(\as^2\right)$    &      6.49  &     1.72  &    4.58   &       7.50   &     1.38    &   3.28   \\
$\ord\left(\as^3\right)$    &     15.35  &    1.03  &    2.49   &     18.67   &     0.35    &   1.58   \\
$\ord\left(\as^4\right)$    &     51.09  &    0.61  &    1.45   &     62.97   &     0.07    &   0.52   \\
$\ord\left(\as^5\right)$    &   217.53  &    0.36  &    0.87   &   269.10   &     0.01    &   0.13   \\ 
$\ord\left(\as^6\right)$    &  1111.56  &    0.22  &    0.52   & 1376.45   &     0.002  &   0.03   \\ \hline
\end{tabular}
\label{tablefix}
\caption{Comparison of the LL contributions to the Higgs production cross section in gluon fusion
expanded in powers of $\alpha_s$ for various choices of the soft and hard scales, and for LP and
NLP, respectively. 
For the naive choice $\mu_s = m_H(1-z)$ the series does not converge at LP and also at NLP,
while higher order contributions decrease rapidly when using the dynamical soft scale $\mu_s = \mu_s^{\rm dyn}$.
Furthermore, we distinguish the case in which $\mu_h^2 = m_H^2$ and $\mu_h^2 = -m_H^2$.}
\end{center}
\end{table}

To evaluate the LL resummed NLP cross section, we find it useful to consider a slight
generalization of our previous result~\eqref{eq:NLPsummedfinalgeneral}, given by
\bea\label{NLPresum} \nn
\Delta^{\rm LL}_{\rm NLP}(z,\mu) 
&=& \bigg[\frac{\beta\big(\as(\mu)\big)}{\as^2(\mu)}
\frac{\as^2(\mu_t)}{\beta\big(\as(\mu_t)\big)}\bigg]^2 
C^2_t(m_t,\mu_t) 
\exp\Big[4 C_A \big(S^{\rm LL}(\mu_h,\mu) - S^{\rm LL}(\mu_s,\mu) \big)\Big] \\
&&\hspace{0.0cm} \times \, \bigg[S_{\rm NLP}\big(m_H(1-z),\mu_s\big) 
- \frac{8 C_A}{\beta_0} \ln \frac{\as(\mu)}{\as(\mu_s)} 
S_{x_0}^{\rm ad} \big(m_H(1-z),\mu_s\big)\bigg],
\eea
where $S_{\rm NLP}\big(m_H(1-z),\mu_s\big) $ denotes 
the initial condition of the evolution equations \eqref{eq:ADMcopy} and \eqref{eq:ADMcopy-kin}
for the sum of NLP soft functions (\ref{eq:NLPfactevolved2})
evaluated at the soft scale,
\be
S_{\rm NLP} \big(m_H (1-z),\mu_s\big) 
 = S_{\rm K}  \big(m_H (1-z),\mu_s\big) 
-8 \int d \omega \, S_{\rm YM} \big(m_H (1-z),\omega,\mu_s\big).
\ee
We consider the two initial conditions
\be
\begin{array}{lll}\label{initialconditions}
{\rm A)} & \quad S_{\rm NLP}\big(m_H(1-z),\mu_s\big) = 0, & \quad S_{x_0}^{\rm ad}\big(m_H(1-z),\mu_s\big) = 1,   \\
{\rm B)} & \quad S_{\rm NLP}\big(m_H(1-z),\mu_s\big) = -4 C_A \frac{\as(\mu_s)}{2\pi} \ln \frac{m_H^2(1-z)^2}{\mu_s^2}, &  
\quad S_{x_0}^{\rm ad}\big(m_H(1-z),\mu_s\big) = 1,   \\
\end{array}
\ee
that are equivalent at LL accuracy.
The first choice was used above and reproduces (\ref{eq:NLPsummedfinalgeneral})
while the second reproduces the logarithmic part of the NLP NLO contribution for 
$\mu_h = \mu_s = \mu$.

To obtain the NLP resummed result, we take $C^2_t(m_t,\mu_t)$ in \eqref{NLPresum} at two loops, see (12) of \cite{Ahrens:2008nc},
and use the three-loop $\beta$-function for $\alpha_s$. This gives $C^2_t(m_t,\mu=m_H)\simeq 1.22$ for $C_t$ evolved
to the factorization scale (given by the product of the first two factors on the right-hand side of \eqref{NLPresum}).
For the LP result, $C_t$ evaluated at the soft scale is required \cite{Ahrens:2008nc}, for which we find $C^2_t(m_t,\mu=\mu_s)\simeq 1.80$.

\begin{table}[t]
\begin{center}
\begin{tabular}{|l|r|r|}
\hline
\multirow{2}{*}{$\sigma$\,(pb)}  &                         \multicolumn{2}{c|}{$\mu_s = \mu_s^{\rm dyn}$}   \\  \cline{2-3}
                                                 &           $\mu^2_h = m_H^2$ &  $\mu^2_h = -m_H^2$ \\ \hline\hline
$\sigma_{\rm LP}^{\rm NNLL}$                               &   24.12    &  28.04   \\  \hline
{\small $\sigma_{\rm LP}^{\rm NNLO}$}                  &   \multicolumn{2}{c|}{28.93}    \\  
{\small $\sigma_{\rm LP}^{{\rm N}^3{\rm LO}}$}                 &   \multicolumn{2}{c|}{29.24}   \\  \hline\hline
$\sigma_{\rm NLP}^{\rm LL}$ (A)                            &     7.18    &  12.76   \\ 
$\sigma_{\rm NLP}^{\rm LL}$ (B)                            &     8.82    &  15.68   \\ \hline
{\small $\sigma_{\rm non\,LP}^{\rm NNLO}$}          &  \multicolumn{2}{c|}{11.90}  \\ 
{\small $\sigma_{\rm non\,LP}^{{\rm N}^3{\rm LO}}$}         &  \multicolumn{2}{c|}{16.27}  \\  \hline\hline
$\sigma_{\rm LP}^{\rm NNLL}+\sigma_{\rm NLP}^{\rm LL}$ (A)        &   31.30    &  40.80   \\ 
$\sigma_{\rm LP}^{\rm NNLL}+\sigma_{\rm NLP}^{\rm LL}$ (B)        &   32.94    &  43.72   \\ \hline
{\small $\sigma^{\rm NNLO}$}                           & \multicolumn{2}{c|}{40.82}    \\  
{\small $\sigma^{{\rm N}^3{\rm LO}}$}                          & \multicolumn{2}{c|}{45.52}  \\  \hline
\end{tabular}
\caption{Resummed Higgs production cross section in gluon fusion at LP with NNLL, and at NLP with LL accuracy. 
For NLP we present the result for the two cases defined in \eqref{initialconditions}. 
For comparison, we also show the fixed order results for gluon fusion at NNLO and N$^3$LO
based on the {\sc iHixs} code \cite{Dulat:2018rbf}. In addition, we distinguish the
LP contribution and the difference between the full result and LP contribution (denoted by non\,LP) for
the fixed-order results.}
\label{tableres1}
\end{center}
\end{table}

In table \ref{tableres1} we present our numerical 
results for the LL resummed NLP cross section within the 
two schemes discussed above, and compare to the NNLL LP
result as well as to  fixed order results at NNLO and N$^3$LO, 
obtained using the {\sc iHixs} code \cite{Dulat:2018rbf}.
We find that the NLP correction is sizeable, and constitutes up to $40\%$ 
of the NNLL LP resummed cross section. Furthermore, we find that the 
resummation of $\pi^2$ enhanced terms, although
contributing formally beyond LL accuracy, is numerically important. 
Its inclusion leads to a combined NNLL LP + LL NLP resummed 
cross section that is comparable to the N$^3$LO 
result \cite{Anastasiou:2016cez,Mistlberger:2018etf}. 

\begin{figure}[t]
\begin{center}
  \includegraphics[width=0.70\textwidth]{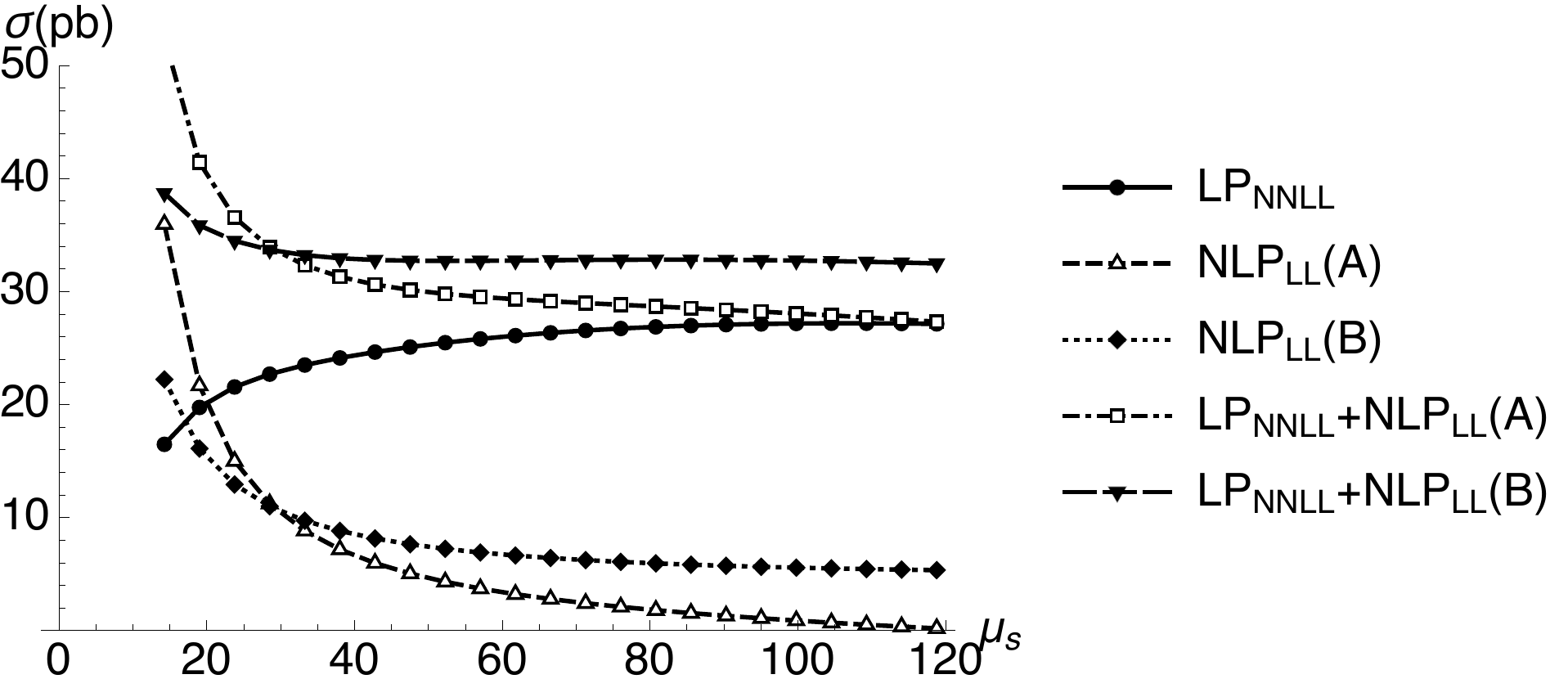}
\end{center} 
\caption{Dependence of the NNLL LP and LL NLP resummed Higgs production 
cross section on the soft scale $\mu_s$,  for $\mu_h^2 = m_H^2$. 
For NLP we present the result for the two cases defined in \eqref{initialconditions}. }
\label{sigmamus}
\end{figure}

\begin{figure}[t]
\begin{center}
  \includegraphics[width=0.70\textwidth]{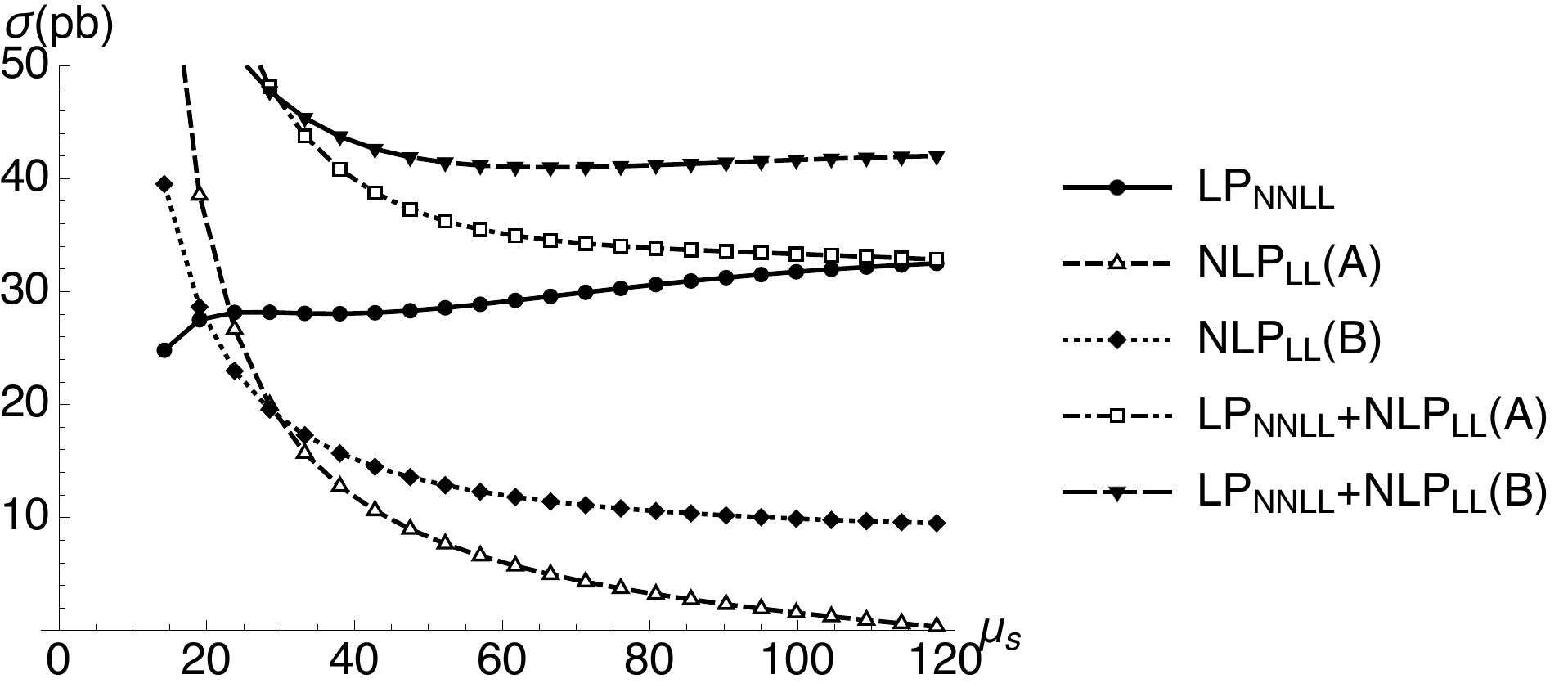}
\end{center} 
\caption{Same as figure~\ref{sigmamus}, however using $\mu_h^2 = -m_H^2$ for the hard scale.}
\label{sigmamusb}
\end{figure}

In figures~\ref{sigmamus} and~\ref{sigmamusb} we study the 
dependence of our resummed result on the soft scale $\mu_s$.
As expected, the sensitivity to $\mu_s$ is larger for the NLP 
corrections, given that only the LLs are available, compared to 
the LP cross section, which is resummed at NNLL
accuracy. These numerical results should therefore be interpreted 
with care. Despite the large sensitivity to $\mu_s$, 
the NLP correction proves to be stable in the region around 
$\mu^{\rm dyn}_s$: the resummed cross 
section obtained for the two initial conditions A and B
overlaps in this region. 

\vspace{0.5cm}
\section{Summary}
\label{sec:summary}
In this work, we resummed all leading logarithmic
corrections to the Higgs production cross section in gluon-gluon fusion,
at next-to-leading power in the threshold expansion employing position-space SCET. 
Our main analytic result is presented in \eqref{eq:NLPsummedfinalgeneral},
and its expansion in powers of $\alpha_s$ is provided
in \eqref{eq:fullexpanded} up to the fifth order. 
Our result proves that, to all
orders in the strong coupling constant, the leading
logarithmic terms at NLP are identical to those obtained
for the threshold expansion of the Drell-Yan process in the
$q\bar q$ channel up to an exchange of the color factor, $C_F\to C_A$.
It also confirms previous conjectures about the structure of the LL 
resummed NLP contribution.

In addition, we explore the relevance of the resummed NLP correction
for the total hadronic Higgs production cross section in
proton-proton collisions at $\sqrt{s}=13$\,TeV.
Similarly to what has been observed at LP, a direct naive summation of
the fixed-order contributions is not possible, as the sum
would not converge in this case. A judicious choice of the soft scale is necessary and to that end 
we employ the soft scale
setting procedure established in the literature for LP
investigations. Even though this method may not be considered
as fully systematic, it is known to provide reasonable results at LP.
We find that this choice provides a series whose subsequent contributions
decrease rapidly also at NLP.

We find that the LL NLP corrections are of the order of 30-40\%
of the NNLL LP resummed cross section (when including only leading power
contributions to the latter). We also observe that the sum of NNLL LP
and LL NLP contributions yields a value that is numerically close to the N$^3$LO 
 result, when including a resummation of $\pi^2$ terms in
both the LP and NLP contributions. The difference between both is of similar size
as the contribution to the resummed result from orders larger than or equal to
$\alpha_s^4$.

As expected, the dependence of the LL NLP result on the
soft scale is sizeable. Therefore, the precise numerical values should be
interpreted with care.
We find that the numerical results are comparable when using
two different initial conditions for the soft function that are
formally equivalent at LL accuracy.  In conclusion, our results indicate that
the NLP correction to the Higgs cross section
is substantial, and it therefore appears desirable
to extend NLP resummation to NLL accuracy, and eventually combine 
the resummed results with fixed-order computations.

\subsubsection*{Acknowledgments} 
We would like to thank Sven-Olaf Moch for a useful discussion.
This work has been supported by the Bundesministerium f\"ur
Bildung and Forschung (BMBF) grant no. 05H18WOCA1, by the
D-ITP consortium, a program of NWO funded by the Dutch Ministry
of Education, Culture and Science, and by Fellini - Fellowship for
Innovation at INFN, funded by the European Union’s Horizon
2020 research programme under the Marie Sk\l{}odowska-Curie
Cofund Action, grant agreement no. 754496.


\begin{thebibliography}{10}

\bibitem{Anastasiou:2015ema}
C.~Anastasiou, C.~Duhr, F.~Dulat, F.~Herzog and B.~Mistlberger, \emph{{Higgs
  Boson Gluon-Fusion Production in QCD at Three Loops}},
  \href{https://doi.org/10.1103/PhysRevLett.114.212001}{\emph{Phys. Rev. Lett.}
  {\bfseries 114} (2015) 212001},
  [\href{https://arxiv.org/abs/1503.06056}{{\ttfamily 1503.06056}}].

\bibitem{Anastasiou:2016cez}
C.~Anastasiou, C.~Duhr, F.~Dulat, E.~Furlan, T.~Gehrmann, F.~Herzog et~al.,
  \emph{{High precision determination of the gluon fusion Higgs boson
  cross-section at the LHC}},
  \href{https://doi.org/10.1007/JHEP05(2016)058}{\emph{JHEP} {\bfseries 05}
  (2016) 058}, [\href{https://arxiv.org/abs/1602.00695}{{\ttfamily
  1602.00695}}].

\bibitem{Mistlberger:2018etf}
B.~Mistlberger, \emph{{Higgs boson production at hadron colliders at N$^{3}$LO
  in QCD}}, \href{https://doi.org/10.1007/JHEP05(2018)028}{\emph{JHEP}
  {\bfseries 05} (2018) 028},
  [\href{https://arxiv.org/abs/1802.00833}{{\ttfamily 1802.00833}}].

\bibitem{Dulat:2018bfe}
F.~Dulat, B.~Mistlberger and A.~Pelloni, \emph{{Precision predictions at
  N$^3$LO for the Higgs boson rapidity distribution at the LHC}},
  \href{https://doi.org/10.1103/PhysRevD.99.034004}{\emph{Phys. Rev.}
  {\bfseries D99} (2019) 034004},
  [\href{https://arxiv.org/abs/1810.09462}{{\ttfamily 1810.09462}}].

\bibitem{Moch:2005ky}
S.~Moch and A.~Vogt, \emph{{Higher-order soft corrections to lepton pair and
  Higgs boson production}},
  \href{https://doi.org/10.1016/j.physletb.2005.09.061}{\emph{Phys. Lett.}
  {\bfseries B631} (2005) 48--57},
  [\href{https://arxiv.org/abs/hep-ph/0508265}{{\ttfamily hep-ph/0508265}}].

\bibitem{Laenen:2005uz}
E.~Laenen and L.~Magnea, \emph{{Threshold resummation for electroweak
  annihilation from DIS data}},
  \href{https://doi.org/10.1016/j.physletb.2005.10.038}{\emph{Phys. Lett.}
  {\bfseries B632} (2006) 270--276},
  [\href{https://arxiv.org/abs/hep-ph/0508284}{{\ttfamily hep-ph/0508284}}].

\bibitem{Idilbi:2005ni}
A.~Idilbi, X.-d. Ji, J.-P. Ma and F.~Yuan, \emph{{Threshold resummation for
  Higgs production in effective field theory}},
  \href{https://doi.org/10.1103/PhysRevD.73.077501}{\emph{Phys. Rev.}
  {\bfseries D73} (2006) 077501},
  [\href{https://arxiv.org/abs/hep-ph/0509294}{{\ttfamily hep-ph/0509294}}].

\bibitem{Idilbi:2006dg}
A.~Idilbi, X.-d. Ji and F.~Yuan, \emph{{Resummation of threshold logarithms in
  effective field theory for DIS, Drell-Yan and Higgs production}},
  \href{https://doi.org/10.1016/j.nuclphysb.2006.07.002}{\emph{Nucl. Phys.}
  {\bfseries B753} (2006) 42--68},
  [\href{https://arxiv.org/abs/hep-ph/0605068}{{\ttfamily hep-ph/0605068}}].

\bibitem{Ahrens:2008nc}
V.~Ahrens, T.~Becher, M.~Neubert and L.~L. Yang, \emph{{Renormalization-Group
  Improved Prediction for Higgs Production at Hadron Colliders}},
  \href{https://doi.org/10.1140/epjc/s10052-009-1030-2}{\emph{Eur. Phys. J.}
  {\bfseries C62} (2009) 333--353},
  [\href{https://arxiv.org/abs/0809.4283}{{\ttfamily 0809.4283}}].

\bibitem{Bonvini:2014joa}
M.~Bonvini and S.~Marzani, \emph{{Resummed Higgs cross section at N$^{3}$LL}},
  \href{https://doi.org/10.1007/JHEP09(2014)007}{\emph{JHEP} {\bfseries 09}
  (2014) 007}, [\href{https://arxiv.org/abs/1405.3654}{{\ttfamily 1405.3654}}].

\bibitem{Beneke:2018gvs}
M.~Beneke, A.~Broggio, M.~Garny, S.~Jaskiewicz, R.~Szafron, L.~Vernazza et~al.,
  \emph{{Leading-logarithmic threshold resummation of the Drell-Yan process at
  next-to-leading power}},
  \href{https://doi.org/10.1007/JHEP03(2019)043}{\emph{JHEP} {\bfseries 03}
  (2019) 043}, [\href{https://arxiv.org/abs/1809.10631}{{\ttfamily
  1809.10631}}].

\bibitem{Sterman:1986aj}
G.~Sterman, \emph{{Summation of Large Corrections to Short Distance Hadronic
  Cross-Sections}}, {\emph{Nucl. Phys.} {\bfseries B281} (1987) 310}.

\bibitem{Catani:1989ne}
S.~Catani and L.~Trentadue, \emph{{Resummation of the QCD Perturbative Series
  for Hard Processes}},
  \href{https://doi.org/10.1016/0550-3213(89)90273-3}{\emph{Nucl. Phys.}
  {\bfseries B327} (1989) 323--352}.

\bibitem{Bahjat-Abbas:2019fqa}
N.~Bahjat-Abbas, D.~Bonocore, J.~Sinninghe~Damst\'{e}, E.~Laenen, L.~Magnea,
  L.~Vernazza et~al., \emph{{Diagrammatic resummation of leading-logarithmic
  threshold effects at next-to-leading power}},
  \href{https://arxiv.org/abs/1905.13710}{{\ttfamily 1905.13710}}.

\bibitem{Laenen:2010uz}
E.~Laenen, L.~Magnea, G.~Stavenga and C.~D. White, \emph{{Next-to-eikonal
  corrections to soft gluon radiation: a diagrammatic approach}},
  \href{https://doi.org/10.1007/JHEP01(2011)141}{\emph{JHEP} {\bfseries 01}
  (2011) 141}, [\href{https://arxiv.org/abs/1010.1860}{{\ttfamily 1010.1860}}].

\bibitem{Bonocore:2015esa}
D.~Bonocore, E.~Laenen, L.~Magnea, S.~Melville, L.~Vernazza and C.~D. White,
  \emph{{A factorization approach to next-to-leading-power threshold
  logarithms}}, \href{https://doi.org/10.1007/JHEP06(2015)008}{\emph{JHEP}
  {\bfseries 06} (2015) 008},
  [\href{https://arxiv.org/abs/1503.05156}{{\ttfamily 1503.05156}}].

\bibitem{DelDuca:2017twk}
V.~Del~Duca, E.~Laenen, L.~Magnea, L.~Vernazza and C.~D. White,
  \emph{{Universality of next-to-leading power threshold effects for colourless
  final states in hadronic collisions}},
  \href{https://doi.org/10.1007/JHEP11(2017)057}{\emph{JHEP} {\bfseries 11}
  (2017) 057}, [\href{https://arxiv.org/abs/1706.04018}{{\ttfamily
  1706.04018}}].

\bibitem{Moult:2018jjd}
I.~Moult, I.~W. Stewart, G.~Vita and H.~X. Zhu, \emph{{First Subleading Power
  Resummation for Event Shapes}},
  \href{https://doi.org/10.1007/JHEP08(2018)013}{\emph{JHEP} {\bfseries 08}
  (2018) 013}, [\href{https://arxiv.org/abs/1804.04665}{{\ttfamily
  1804.04665}}].

\bibitem{Beneke:2017ztn}
M.~Beneke, M.~Garny, R.~Szafron and J.~Wang, \emph{{Anomalous dimension of
  subleading-power N-jet operators}},
  \href{https://doi.org/10.1007/JHEP03(2018)001}{\emph{JHEP} {\bfseries 03}
  (2018) 001}, [\href{https://arxiv.org/abs/1712.04416}{{\ttfamily
  1712.04416}}].

\bibitem{Beneke:2018rbh}
M.~Beneke, M.~Garny, R.~Szafron and J.~Wang, \emph{{Anomalous dimension of
  subleading-power $N$-jet operators. Part II}},
  \href{https://doi.org/10.1007/JHEP11(2018)112}{\emph{JHEP} {\bfseries 11}
  (2018) 112}, [\href{https://arxiv.org/abs/1808.04742}{{\ttfamily
  1808.04742}}].

\bibitem{Beneke:2019kgv}
M.~Beneke, M.~Garny, R.~Szafron and J.~Wang, \emph{{Violation of the
  Kluberg-Stern-Zuber theorem in SCET}},
  \href{https://doi.org/10.1007/JHEP09(2019)101}{\emph{JHEP} {\bfseries 09}
  (2019) 101}, [\href{https://arxiv.org/abs/1907.05463}{{\ttfamily
  1907.05463}}].

\bibitem{Marcantonini:2008qn}
C.~Marcantonini and I.~W. Stewart, \emph{{Reparameterization Invariant
  Collinear Operators}},
  \href{https://doi.org/10.1103/PhysRevD.79.065028}{\emph{Phys. Rev.}
  {\bfseries D79} (2009) 065028},
  [\href{https://arxiv.org/abs/0809.1093}{{\ttfamily 0809.1093}}].

\bibitem{Kolodrubetz:2016uim}
D.~W. Kolodrubetz, I.~Moult and I.~W. Stewart, \emph{{Building Blocks for
  Subleading Helicity Operators}},
  \href{https://doi.org/10.1007/JHEP05(2016)139}{\emph{JHEP} {\bfseries 05}
  (2016) 139}, [\href{https://arxiv.org/abs/1601.02607}{{\ttfamily
  1601.02607}}].

\bibitem{Feige:2017zci}
I.~Feige, D.~W. Kolodrubetz, I.~Moult and I.~W. Stewart, \emph{{A Complete
  Basis of Helicity Operators for Subleading Factorization}},
  \href{https://doi.org/10.1007/JHEP11(2017)142}{\emph{JHEP} {\bfseries 11}
  (2017) 142}, [\href{https://arxiv.org/abs/1703.03411}{{\ttfamily
  1703.03411}}].

\bibitem{Moult:2017rpl}
I.~Moult, I.~W. Stewart and G.~Vita, \emph{{A subleading operator basis and
  matching for $gg\to H$}},
  \href{https://doi.org/10.1007/JHEP07(2017)067}{\emph{JHEP} {\bfseries 07}
  (2017) 067}, [\href{https://arxiv.org/abs/1703.03408}{{\ttfamily
  1703.03408}}].

\bibitem{BBJVunpublished}
M.~Beneke, A.~Broggio, S.~Jaskiewicz and L.~Vernazza{\emph{\hspace*{0.00cm}{\rm
  , in preparation}\hspace*{-0.09cm}} }.

\bibitem{Beneke:2002ni}
M.~Beneke and T.~Feldmann, \emph{{Multipole expanded soft collinear effective
  theory with non-abelian gauge symmetry}},
  \href{https://doi.org/10.1016/S0370-2693(02)03204-5}{\emph{Phys. Lett.}
  {\bfseries B553} (2003) 267--276},
  [\href{https://arxiv.org/abs/hep-ph/0211358}{{\ttfamily hep-ph/0211358}}].

\bibitem{Bauer:2001yt}
C.~W. Bauer, D.~Pirjol and I.~W. Stewart, \emph{{Soft collinear factorization
  in effective field theory}},
  \href{https://doi.org/10.1103/PhysRevD.65.054022}{\emph{Phys. Rev.}
  {\bfseries D65} (2002) 054022},
  [\href{https://arxiv.org/abs/hep-ph/0109045}{{\ttfamily hep-ph/0109045}}].

\bibitem{Bell:2010mg}
G.~Bell, M.~Beneke, T.~Huber and X.-Q. Li, \emph{{Heavy-to-light currents at
  NNLO in SCET and semi-inclusive $\bar{B} \to X_s l^+ l^-$ decay}},
  \href{https://doi.org/10.1016/j.nuclphysb.2010.09.022}{\emph{Nucl. Phys.}
  {\bfseries B843} (2011) 143--176},
  [\href{https://arxiv.org/abs/1007.3758}{{\ttfamily 1007.3758}}].

\bibitem{Anastasiou:2014lda}
C.~Anastasiou, C.~Duhr, F.~Dulat, E.~Furlan, T.~Gehrmann, F.~Herzog et~al.,
  \emph{{Higgs Boson Gluon-Fusion Production Beyond Threshold in N$^{3}LO$
  QCD}}, \href{https://doi.org/10.1007/JHEP03(2015)091}{\emph{JHEP} {\bfseries
  03} (2015) 091}, [\href{https://arxiv.org/abs/1411.3584}{{\ttfamily
  1411.3584}}].

\bibitem{deFlorian:2014vta}
D.~de~Florian, J.~Mazzitelli, S.~Moch and A.~Vogt, \emph{{Approximate N$^{3}$LO
  Higgs-boson production cross section using physical-kernel constraints}},
  \href{https://doi.org/10.1007/JHEP10(2014)176}{\emph{JHEP} {\bfseries 10}
  (2014) 176}, [\href{https://arxiv.org/abs/1408.6277}{{\ttfamily 1408.6277}}].

\bibitem{Kramer:1996iq}
M.~Kramer, E.~Laenen and M.~Spira, \emph{{Soft gluon radiation in Higgs boson
  production at the LHC}},
  \href{https://doi.org/10.1016/S0550-3213(97)00679-2}{\emph{Nucl. Phys.}
  {\bfseries B511} (1998) 523--549},
  [\href{https://arxiv.org/abs/hep-ph/9611272}{{\ttfamily hep-ph/9611272}}].

\bibitem{Kidonakis:2007ww}
N.~Kidonakis, \emph{{Collinear and soft gluon corrections to Higgs production
  at NNNLO}}, \href{https://doi.org/10.1103/PhysRevD.77.053008}{\emph{Phys.
  Rev.} {\bfseries D77} (2008) 053008},
  [\href{https://arxiv.org/abs/0711.0142}{{\ttfamily 0711.0142}}].

\bibitem{Buckley:2014ana}
A.~Buckley, J.~Ferrando, S.~Lloyd, K.~Nordstr{\"o}m, B.~Page, M.~R{\"u}fenacht
  et~al., \emph{{LHAPDF6: parton density access in the LHC precision era}},
  \href{https://doi.org/10.1140/epjc/s10052-015-3318-8}{\emph{Eur. Phys. J.}
  {\bfseries C75} (2015) 132},
  [\href{https://arxiv.org/abs/1412.7420}{{\ttfamily 1412.7420}}].

\bibitem{Butterworth:2015oua}
J.~Butterworth et~al., \emph{{PDF4LHC recommendations for LHC Run II}},
  \href{https://doi.org/10.1088/0954-3899/43/2/023001}{\emph{J. Phys.}
  {\bfseries G43} (2016) 023001},
  [\href{https://arxiv.org/abs/1510.03865}{{\ttfamily 1510.03865}}].

\bibitem{Dulat:2015mca}
S.~Dulat, T.-J. Hou, J.~Gao, M.~Guzzi, J.~Huston, P.~Nadolsky et~al.,
  \emph{{New parton distribution functions from a global analysis of quantum
  chromodynamics}},
  \href{https://doi.org/10.1103/PhysRevD.93.033006}{\emph{Phys. Rev.}
  {\bfseries D93} (2016) 033006},
  [\href{https://arxiv.org/abs/1506.07443}{{\ttfamily 1506.07443}}].

\bibitem{Harland-Lang:2014zoa}
L.~A. Harland-Lang, A.~D. Martin, P.~Motylinski and R.~S. Thorne, \emph{{Parton
  distributions in the LHC era: MMHT 2014 PDFs}},
  \href{https://doi.org/10.1140/epjc/s10052-015-3397-6}{\emph{Eur. Phys. J.}
  {\bfseries C75} (2015) 204},
  [\href{https://arxiv.org/abs/1412.3989}{{\ttfamily 1412.3989}}].

\bibitem{Ball:2014uwa}
{\scshape NNPDF} collaboration, R.~D. Ball et~al., \emph{{Parton distributions
  for the LHC Run II}},
  \href{https://doi.org/10.1007/JHEP04(2015)040}{\emph{JHEP} {\bfseries 04}
  (2015) 040}, [\href{https://arxiv.org/abs/1410.8849}{{\ttfamily 1410.8849}}].

\bibitem{Becher:2007ty}
T.~Becher, M.~Neubert and G.~Xu, \emph{{Dynamical Threshold Enhancement and
  Resummation in Drell- Yan Production}},
  \href{https://doi.org/10.1088/1126-6708/2008/07/030}{\emph{JHEP} {\bfseries
  07} (2008) 030}, [\href{https://arxiv.org/abs/0710.0680}{{\ttfamily
  0710.0680}}].

\bibitem{Sterman:2013nya}
G.~Sterman and M.~Zeng, \emph{{Quantifying Comparisons of Threshold
  Resummations}}, \href{https://doi.org/10.1007/JHEP05(2014)132}{\emph{JHEP}
  {\bfseries 05} (2014) 132},
  [\href{https://arxiv.org/abs/1312.5397}{{\ttfamily 1312.5397}}].

\bibitem{Catani:1996yz}
S.~Catani, M.~L. Mangano, P.~Nason and L.~Trentadue, \emph{{The Resummation of
  soft gluons in hadronic collisions}},
  \href{https://doi.org/10.1016/0550-3213(96)00399-9}{\emph{Nucl. Phys.}
  {\bfseries B478} (1996) 273--310},
  [\href{https://arxiv.org/abs/hep-ph/9604351}{{\ttfamily hep-ph/9604351}}].

\bibitem{Moch:2005id}
S.~Moch, J.~A.~M. Vermaseren and A.~Vogt, \emph{{The Quark form-factor at
  higher orders}},
  \href{https://doi.org/10.1088/1126-6708/2005/08/049}{\emph{JHEP} {\bfseries
  08} (2005) 049}, [\href{https://arxiv.org/abs/hep-ph/0507039}{{\ttfamily
  hep-ph/0507039}}].

\bibitem{Dulat:2018rbf}
F.~Dulat, A.~Lazopoulos and B.~Mistlberger, \emph{{iHixs 2 -- Inclusive Higgs
  cross sections}},
  \href{https://doi.org/10.1016/j.cpc.2018.06.025}{\emph{Comput. Phys. Commun.}
  {\bfseries 233} (2018) 243--260},
  [\href{https://arxiv.org/abs/1802.00827}{{\ttfamily 1802.00827}}].

\end{thebibliography}

\providecommand{\href}[2]{#2}\begingroup\raggedright\endgroup


\end{document}